\documentclass[twocolumn,showpacs,preprintnumbers,amsmath,amssymb]{revtex4}

\usepackage{graphics}

\usepackage{dcolumn}
\usepackage{bm}

\begin{document}

\draft


\title{Two freeze-out model for the hadrons produced in the Relativistic Heavy-Ion Collisions}
\author{Suk Choi , Kang Seog Lee\footnote{Correspond to kslee@chonnam.ac.kr}}
\address{Department of Physics, Chonnam National University,
Gwangju 500-757, Korea }

\date{July 13, 2011, revised on Oct. 17, 2011}

\begin{abstract}

An expanding fireball model with two freeze-outs, which assumes that the chemical freeze-out occurs earlier at higher temperature and the thermal freeze-out occurs later at lower temperature, is developed and successfully applied to fit simultaneously the hadron ratios, transverse momentum spectra of measured hadrons and rapidity distribution of charged hadrons measured in Au+Au collision at 200 A$\cdot$GeV by STAR, PHENIX and PHOBOS collaborations. The quality of all the fittings is very good and the resulting parameters are in agreement with published values.

\end{abstract}


\maketitle

%

\section{INTRODUCTION}

The hadron production in relativistic heavy-ion collisions have been extensively studied:
The multiplicities of various hadrons or the ratios among them are nicely fitted using the statistical distributions with a few parameters such as the temperature, $T_{ch}$, the baryon chemical potential, $\mu_{B,ch}$, the strangeness chemical potential, $\mu_{s,ch}$ and the strangeness fugacity, $\gamma_s$\cite{braun1,manninen,rafelski,pbm,becat,star1}. The temperature thus obtained at RHIC energies is so high and close to the phase transition temperature to quark-gluon plasma and suggests that the hadrons are chemically frozen out just after the hadronization. The value of $\gamma_s$ close to one suggests that the strangeness is nearly equilibrated.
The slopes of the transverse momentum spectra below 2GeV/c of various hadrons are also nicely fitted from an expanding fireball model with a single set of parameters such as the temperature $T_{th}$, the baryon chemical potential $\mu_{B,th}$, the strangeness chemical potential $\mu_{s,th}$, and the transverse expansion velocity $\beta$\cite{star2}. The success of the thermal analysis with the Lorentz boosted thermal distribution is regarded as the evidence of the radial expansion.

However, the temperatures of the two analysis are different, {\it i.e.} $T_{ch} \ne T_{th}$ and thus one cannot fit both the magnitudes and slopes of the transverse momentum spectra of various hadrons. In other words, when one fits the transverse momentum spectra normalization constant has to to be adjusted separately for each hadron species. The difference in the two freeze-out temperatures are interpreted as the chemical freeze-out occurs earlier at high temperature where the inelastic collisions of the type A+B $\rightleftharpoons$ C+D becomes less frequent and the numbers of each hadron species are no more changing and thus kept fixed, while during the continuing expansion of the system, elastic collisions among the same species maintains the thermal equilibrium and lowers the temperature until the thermal freeze-out.

In the hydrodynamic calculations, the difference of the two freeze-outs has to be taken into consideration when one wants to compare the results of calculation with the experimental data. Such models reported so far are the hybrid models\cite{bass,hirano,song} and the partial chemical equilibrium model(PCE)\cite{gerber,teaney,hirano,kolb,huovinen}.
In the hybrid models or the hydro+transoport models, the expansion is described by the hydrodynamics, starting from an initial state of QGP  and at $T_c$ the equation of state is changed to the hadronic EOS. At a certain switching temperature in the hadronic phase, $T_{sw}$, hadrons are generated by the Monte Carlo method and the subsequent interaction among hadrons is described using microscopic hadronic transport models such as UrQMD\cite{bass,song} or JAM\cite{hirano}.
In the PCE model, after the chemical freeze-out the numbers of hadrons whose lifetime is longer than the characteristic time($\sim$ $10$ fm/c) are kept fixed, while the reactions involving hadrons with shorter lifetime(e.g. $\Delta \leftrightarrows p + \pi$) are still in chemical equilibrium. In this case one has to solve large number of coupled equations to get the values of chemical potentials.

In the blast-wave model the radial expansion is one of the essential ingredients in fitting the transverse momentum spectra with the Lorentz boosted thermal distribution which does not change the total number.
Thus in principle, if one uses the Lorentz transformed distribution functions in the chemical analysis, the result is expected to be the same as that without the expansion. Then it seems natural to make an expanding fireball model in which the chemical freeze-out occurs earlier at higher temperature  and then thermal freeze-out occurs at lower temperature, while in between the two temperatures the number of each hadron species is fixed. Even though the concept of the two freeze-outs are already adopted in the hydrodynamic calculations as discussed above, there is no blast-wave type calculations reported until now which fits both the ratios and the particle spectra of various hadrons in a single model.  Then the range of parameters which fit the data with similar quality in the chemical and thermal analysis may be pinned down to a single set of parameters.

In this work we report on a simultaneous blast wave-type analysis of both the hadron ratios and transverse momentum spectra measured at RHIC assuming that the chemical freeze-out occurs earlier at higher temperature and the number of each hadron species is kept fixed until the thermal freeze-out when the particles escape from the system to reach the detectors. At the temperature higher than the chemical freeze-out temperature $T_{ch}$, the chemical potentials for hadrons are the algebraic sums of the baryon and strangeness chemical potentials. The ratios among various hadrons are fitted  using the blast-wave equation for the transverse momentum spectra\cite{dobler} by integrating over $p_t$. The contribution from the decay of high lying resonances are also taken into account by integrating over $p_t$ the resonance contribution to the transverse momentum spectra\cite{sollfrank}.

Below $T_{ch}$, the chemical potential of each hadron species is no more the linear sum of the baryon and the strangeness chemical potentials but it should be numerically found to guarantee the fixed thermal number of hadrons. Using the chemical potential for each hadron species thus obtained one can fit the momentum spectra as done by Dobler et al\cite{dobler} which is the most sophisticated version of blast-wave model.

In this way we have successfully fitted both the ratios among hadrons, transverse momentum spectra of pions, kaons and protons, and the rapidity spectra of the charged hadrons.

One difference of this calculation with the PCE model is in the treatment of the short lived resonances. In this work they are treated as stable particles until the thermal freeze-out and their decay  has been considered after the freeze-out. This approximation causes a small errors but reduces the computation significantly since the coupled equations for the chemical potentials now reduces to independent equations.

\section{ Blast-wave Model with Two Freeze-outs }

Starting point of particles emitted in the heavy-ion collisions is the Cooper-Frye formula\cite{cooper} for the particle invariant spectrum,
\begin{equation}
E\frac{d^3N}{d^3p} = \frac{g}{(2\pi)^3} \int_{\Sigma_f} p^\mu
d\sigma_\mu (x) f(x,p),
\end{equation}
where
\begin{equation}
 f(x,p) = \exp(-\frac{p_\nu u_\nu (x) - \mu }{T}),
\end{equation}
Dobler {\it et al.}\cite{dobler} have assumed the boost invariance in the longitudinal direction\cite{bjorken}
 and the ellipsoidal geometry with azimuthal symmetry. Thus the expansion velocity at $(t,z)$ is $v_L =z/t$ and the corresponding rapidity is $\eta = \tanh^{-1} z/t$ whose value at the surface is $\eta_{max}$. The transverse rapidity at the radial distance $r$ from the center is $\rho(r)=\rho_0 (r/r_{max})^\alpha$, where
$\alpha=1$ is chosen as it gives better fit to the particle spectra. The transverse radius may depend on the longitudinal position, $ r_{max}(\eta) =R_0 \sqrt{1-\frac{\eta^2}{\eta_{max}^2}}$ but in this work it is kept constant. Then one gets for the hadron spectrum
\begin{eqnarray}
\frac{d^2 N_i^{th}}{m_T dm_T dy}&=&\frac{d_i V_{eff}}{(2\pi)^2}\gamma_s \int^{\eta_{max}} _{-\eta_{max}} d\eta \int^{r_{max} (\eta)} _0 rdr m_T \cosh(y-\eta) \nonumber \\
 & & \times \exp{(-\frac{m_T \cosh(y-\eta)\cosh{\rho} -\mu_i}{T})}\, I_0 (\frac{p_T \sinh{\rho}}{T})
 \label{spec}
\end{eqnarray}
where $\gamma_s$ is the strangeness fugacity.

It should be emphasized that the chemical potential, $\mu_i$ is different above and below the chemical freeze-out temperature, $T_{ch}$. Until the chemical freeze-out (for $T\geq T_{ch} $), $\mu_i$ is the algebraic sum of the baryon and the strangeness chemical potentials:
\begin{equation}
\mu_i = (n_q -n_{\bar{q}} ) \mu_q +(n_s -n_{\bar{s}} ) \mu_s .
 \label{mu_ch}
\end{equation}
Below $T_{ch}$, however, Eq.~(\ref{mu_ch}) does not hold and $\mu_i$ should be obtained from the number of
thermal hadrons of type $i$, which is fixed at the chemical freeze-out as will be discussed later.

For the number of thermal hadrons, $N_i^{th}$, one integrates Eq.~(\ref{spec}) over $m_T dm_T dy$.
\begin{equation}
N_i^{th} = \int m_T dm_T dy (\frac{d^2 N_i^{th}}{m_T dm_T dy})
\label{number}
\end{equation}
The total number of hadrons measured is the sum of the thermal ones and the decayed ones from other hadrons or resonances. The resonance contribution to the hadron momentum spectra is nicely treated by Sollfrank\cite{sollfrank}, and to get the number of hadrons from the resonance decay one should integrate the spectrum over the transverse mass and the rapidity. Hence one get for the total number of particle species $i$ as
\begin{equation}
N_i = N_i^{th} +N_i^{res}
\end{equation}

Using the above equation, one can fit the ratios among hadrons to get the temperature $T_{ch}$, the chemical potentials $\mu_B$ and $\mu_s$.
The parameters, $\eta_{max}$, $\rho_0$, $R_0$ and the overall factor $V_{eff}$ do net affect the ratios as expected and their values at the chemical freeze-out cannot be determined.
However, $\rho_0$ and $\eta_{max}$ play crucial roles in determining the slope of the transverse mass spectra and width of the rapidity distribution of the total charged hadrons, correspondingly.
It should be noted that the ranges of the rapidity integration in Eq.~({\ref{number}}), which may differ from particle species to another, should match the experimental cut-off. Due to the lack of detailed information we have integrated from -3 to +3 for all the particles and this may be one source of errors in this calculation. Another source of error is the lower limit of integration over the transverse momentum.
We have chosen it to be zero but it is not. In the future studies it should match the experimental value.

Depending on the data processing procedure whether the weak decay contribution from resonances is included or not, weak decay contribution should be correspondingly added or not in calculating the resonance contribution. For example, the PHENIX data for protons and anti-protons were already subtracted by the weak decay contribution from the lambda and anti-lambda particles whereas for pions, they are included. Thus we have fitted PHENIX data separately from the STAR data. In fitting the PHENIX ratios, we subtracted the pions from the decay of lamda's as subtracting out the contribution to protons and including it to pions is very cumbersome and the contribution to pions is negligible.

The fitted parameters are listed in the TABLE 1. The first low is the fitted parameters from STAR data and the second one is for PHENIX data. The difference in the temperatures is less than 1 MeV while chemical potentials differ slightly. The difference may come from the experimental cuts for rapidity and transverse momentum, especially the lower limit of $p_t$, for each particle species.
 The chemical freeze-out temperature of 173Mev and chemical potentials are in agreement with the values from other statistical models(160-170 MeV). The columns which is empty in this table are irrelevant parameters as mentioned earlier.
In the STAR analysis we have omitted $\Xi/\pi$ ratio since it cannot be fitted well in this calculation.
The value for $\chi^2 /n$ in TABLE 1 is obtained without $\Xi/\pi$ ratio while when we include it the value for $\chi^2 /n$ increases to 2.85. The $\phi/K^-$ ratio by PHENIX collaboration cannot fitted in this calculation but the value is even twice smaller compared to STAR measurement. $\chi^2 /n$ value in TABLE 1 for PHENIX data is obtained without this ratio.
 In Fig. 1 and Fig. 2, the results of our chemical analysis are shown and compared with the STAR data and PHENIX data, respectively. Fig. 1 shows that the fitting is very good except for the $\Xi/\pi$ ratio and in Fig. 2 fitting is extremely good for all the ratios except for $\phi/K$ ratio, which differs from STAR data. Our fitting of this ratio to STAR data is rather good.

Below the chemical freeze-out, the number of thermal hadrons of any type $i$ is now fixed and known numerically as the number of inelastic collisions is not sufficient enough. Whereas elastic collisions are still abundant to make the local thermal equilibrium and thus during the continuing expansion of the system, the temperature drops until the number of elastic collisions is not enough to maintain even the thermal equilibrium. When the system breaks up the hadrons have the momentum distribution at this point, the so called thermal freeze-out.
Now the transverse mass spectra and the rapidity distribution of measured hadrons are given from  Eq.~(\ref{spec}) by integrating over either the rapidity or the transverse momentum. The resonance contributions\cite{dobler} should be correctly added. Explicitly, the transverse momentum spectrum is given as
\begin{equation}
 \frac{d N_i}{m_T dm_T } = \int (\frac{d^2 N_i^{th}}{m_T dm_T dy}) dy + \mbox{(res. contr.)}
 \end{equation}
 and the rapidity distribution is given as
\begin{equation}
 \frac{d N_i}{dy} = \int (\frac{d^2 N_i^{th}}{m_T dm_T dy}) m_T dm_T + \mbox{(res. contr.)}
 \end{equation}

It should be stressed again that Eq.~(\ref{mu_ch}) does not hold in above equations since the chemical equilibrium is already broken. Rather, $\mu_i$ can be found from the following equation such that it gives the right number of the thermal particles, $N_i^{th}$, which is known from the chemical freeze-out.
\begin{equation}
N_i^{th} = \int\int m_T dm_T dy \frac{d^2 N_i^{th}}{m_T dm_T dy} (T,\mu_i ,\eta_{max} ,\rho_0, R_0 )
\end{equation}
Instead of using the above equation for the number to calculate $\mu_i$, one can determine it relative to the pion chemical potential from the fixed ratios relative to pions, $R_{i\pi} = N_i^{th}/N_{\pi}^{th}$. Then $\mu_i $ can be obtained from

\begin{equation}
\mu_i = \mu_{\pi} +T \ln{[R_{i\pi} \frac{\int\int m_T dm_T dy (d^2N_i ^{'} /m_T dm_T dy)}
 {\int\int m_T dm_T dy (d^2N_{\pi} ^{'} /m_T dm_T dy)}  ]}
\end{equation}
where the $'$ denotes that $\exp{(\mu_i /T)}$ is absent in this equation compared to Eq.~(\ref{spec}). Now $V_{eff} \cdot e^{\mu_\pi /T}$ acts as an overall constant for all the particle spectra.

The resulting thermal freeze-out parameters are listed in the TABLE 1 together with the chemical freeze-out values. Thus obtained transverse momentum spectra of the charged pions, kaons and protons are shown together with the experimental data by the STAR collaborations in Fig. 3. The quality of fitting is rather good showing that the blast-wave model works very well for the transverse mass spectra in heavy-ion collisions.

The parameter $\eta_{max}$ which is the rapidity of the fireball at the longitudinal surface affects very little on the transverse momentum spectra. However, it plays crucial role in determining the width of rapidity distribution and thus we have fitted $\eta_{max}$ to the pseudo-rapidity distribution of the charged hadrons measured by the PHOBOS collaboration by fixing all other parameters. The fitted value is  5.0 as listed in the TABLE 1.
In Fig. 4, we have drawn the rapidity distribution of charged hadrons together with the PHOBOS data. The shape especially the width of the rapidity distribution fit well to the data but the magnitude in the midrapidity region overpredicts the data. This discrepancy may come from the low $p_T$ cut-off in experiments while we have integrated over $p_T$ from zero. The wiggle near $y \sim 0$ region is due to the errors in calculating the resonance contribution and will vanish if we make y-bin in our calculation smaller.
It should be pointed that the statistical model cannot fit the large width of the in distribution and the present model is very successful in fitting the various aspects of the hadron production except for the correlations. As mentioned earlier $V_{eff} \cdot e^{\mu_\pi /T}$ is an overall constant now. For $V_{eff}$ we get 15597, and $\mu_{\pi} = 115$ MeV.


\section{ Summary and conclusion }

Assuming that the chemical freeze-out occurs earlier at higher temperature and thermal freeze-out occurs later at lower temperature, a blast-wave model with resonance contribution have been successfully applied to fit the ratios among hadrons, the transverse momentum spectra of pions, kaons and protons by STAR and PHENIX collaborations and the rapidity distribution of all the charged hadrons by PHOBOS collaborations measured at RHIC in Au+Au collisions at 200 A$\cdot$GeV. This is a consistent way of describing all the features of hadron production in heavy-ion collisions except for correlations. The result of fitting is very good except for $\Xi / \pi$ ratio from STAR collaboration and $\phi/pi$ ratio from PHENIX collaboration. The  parameters thus obtained are in agreement with the published values obtained separately either by the chemical analysis or the thermal analysis. It should be stressed that in order to analyze the particle production detailed information on the experimental cuts for $p_T$ and $y$ for each particles is needed.
Further studies with LHC data is underway.

\begin{table}
\caption{Fitted values for each parameter: The firs and second lows summarize the chemical freeze-out values of the data by STAR and PHENIX collaborations. Empty columns are irrelevant parameters. }
\begin{tabular} {|c|c|c|c|c|c|c|c|}\hline
    & T &$\mu_B$  & $\mu_s$ & $\gamma_s$& $\eta_{max}$ & $\rho_0$ &  $\chi^2$ /n \\
   &(MeV)  &(MeV) & (MeV)  && & & \\   \hline

STAR & 173.4  & 18.5 & 7.9 & 0.986 &  &   & 1.4 \\ \hline
PHENIX & 173.9  & 26.4 & 6.0 & 1.01 &  &   & 0.12 \\ \hline
thermal f.o. & 121.1 &  &  & 0.986 & 5.0 & 1.03  & 5.3  \\ \hline

\end{tabular}
\end{table}

 \acknowledgments

{K.S. Lee thank U. W. Heinz for suggesting the problem in this work.
This research is supported by Basic Science Research Program through the National Research Foundation of Korea(NRF) funded by the Ministry of Education, Science and Technology(2011-0010433).}


\begin{figure}

\begin{center}
\resizebox{120mm}{!}{\includegraphics{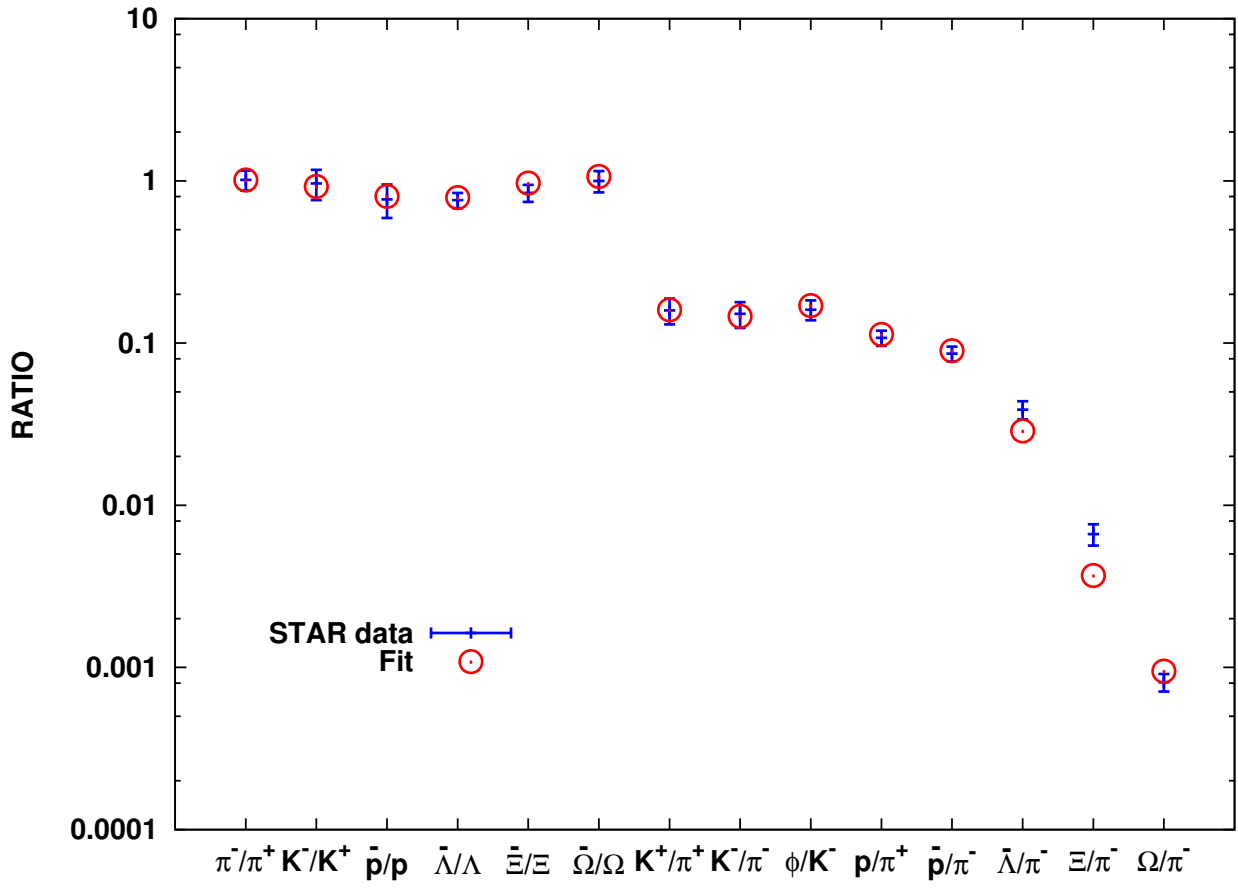}}
\end{center}
\caption{Ratios of various hadrons at RHIC by STAR collaboration\cite{star1,star2,star3,star4,star5}. Fit is quite nice except for $\Xi / \pi$ ratio.}
\label{fig:ratio}
\end{figure}

\begin{figure}

\begin{center}
\resizebox{120mm}{!}{\includegraphics{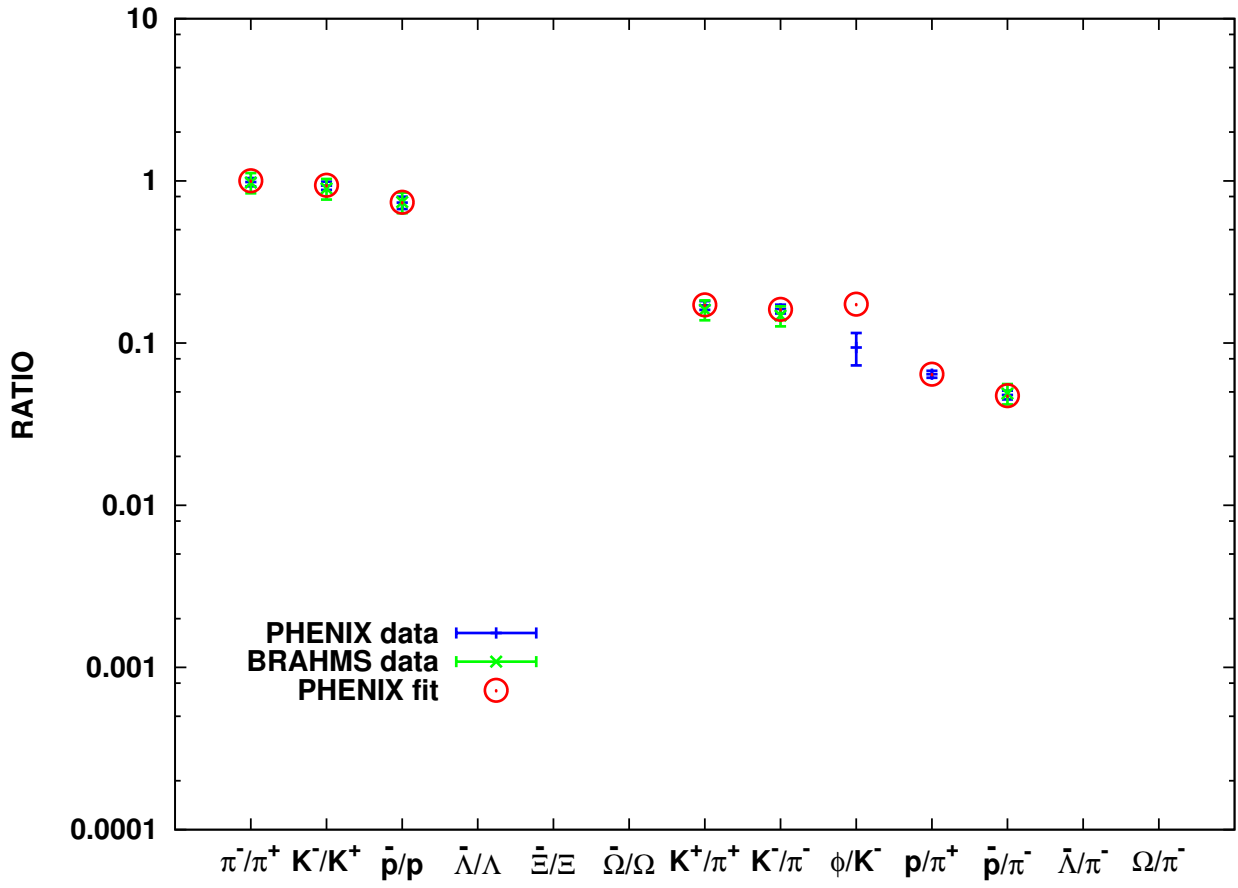}}
\end{center}
\caption{Ratios of various hadrons by PHENIX collaboration\cite{phenix}. Calculated value of $\phi /K^-$ is roughly twice of the measured value.}
\label{fig:ratio_phenix}
\end{figure}

\begin{figure}

\begin{center}
\resizebox{120mm}{!}{\includegraphics{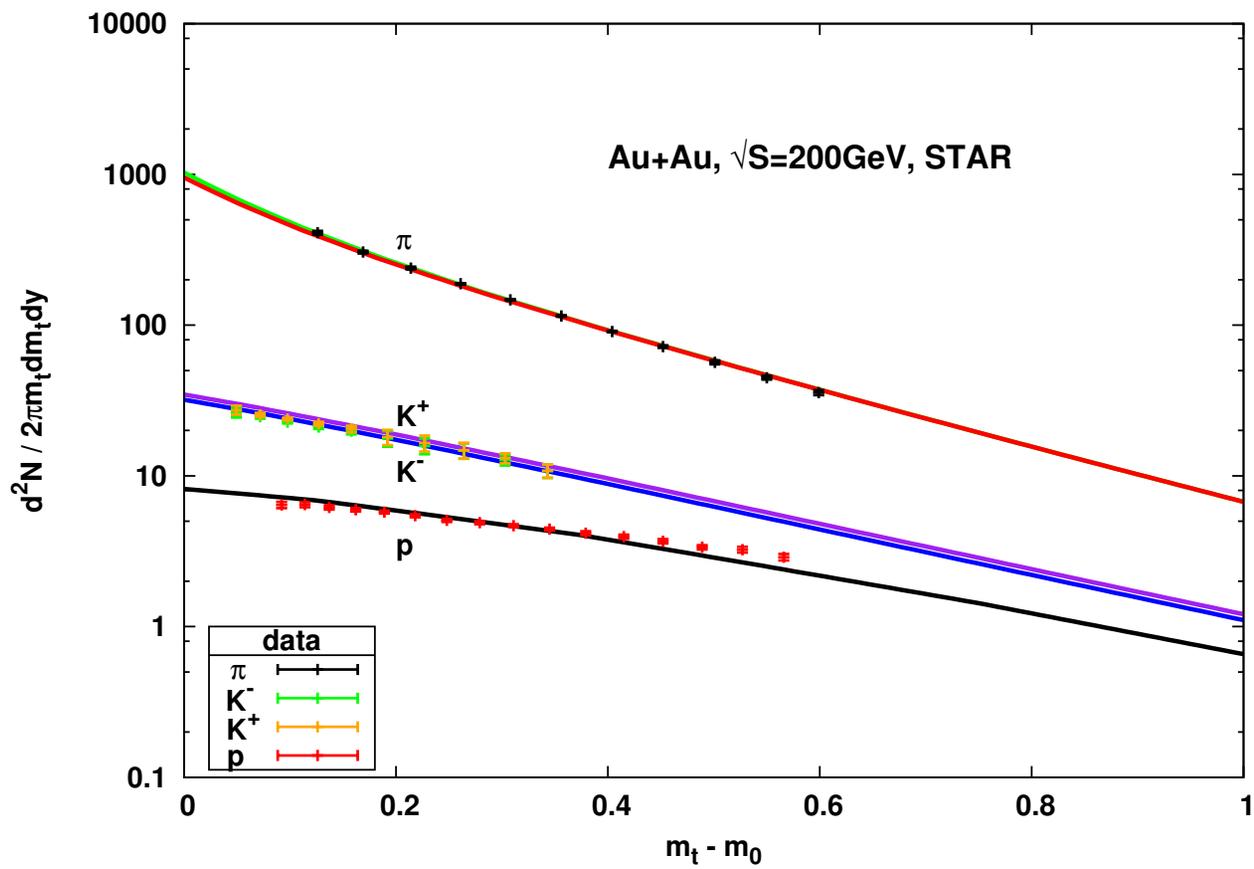}}
\end{center}
\caption{Fit of the transverse mass spectra of hadrons by STAR collaboration\cite{star2} }
\label{fig:mt}
\end{figure}

\begin{figure}

\begin{center}
\resizebox{120mm}{!}{\includegraphics{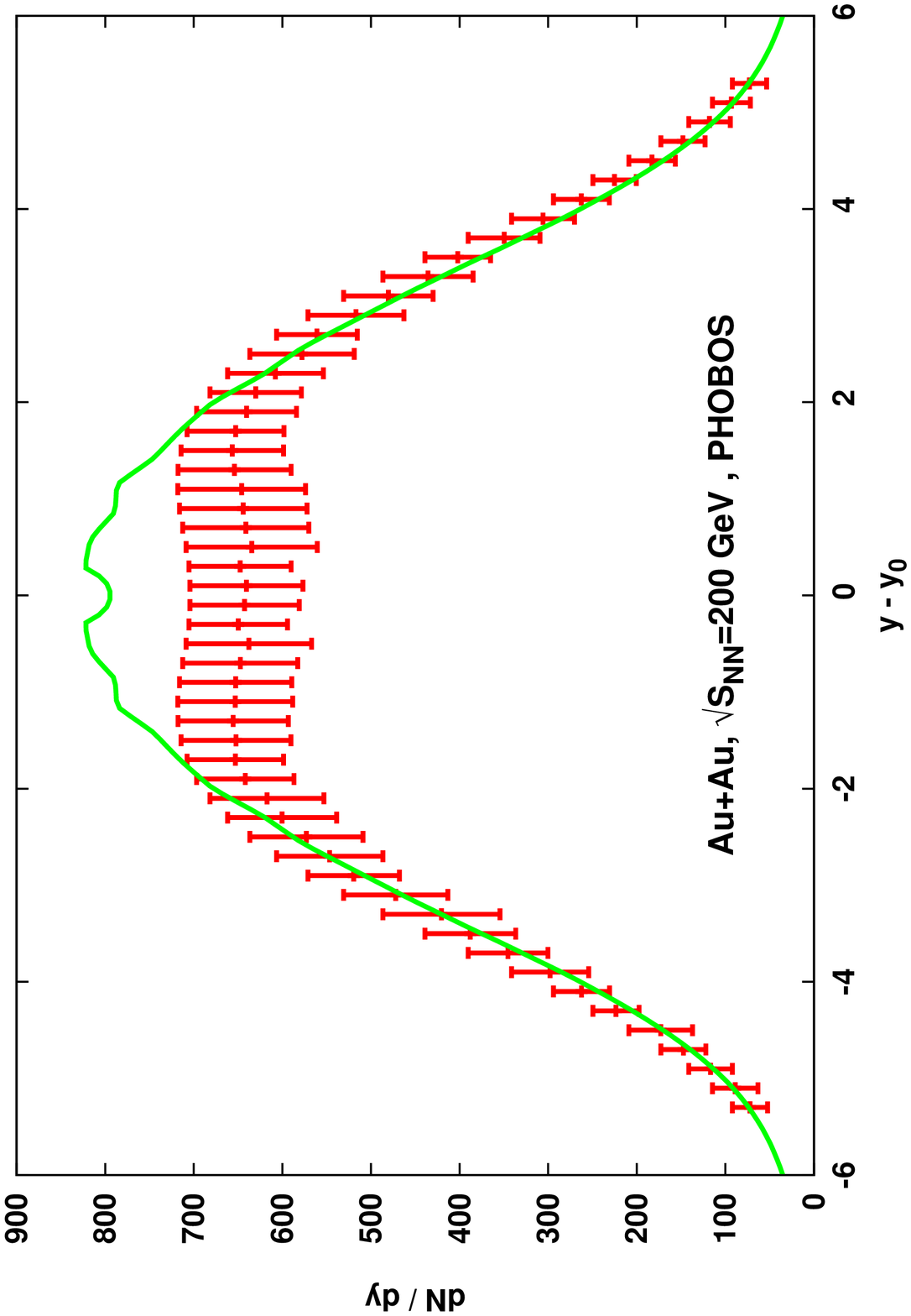}}
\end{center}
\caption{ Comparison of the rapidity distribution of charged hadrons by PHOBOS collaboration\cite{phobos} and calculation. Only $eta_{max}$ and overall constant are fitted and all other parameters are same as chemical analysis. }
\label{fig:rap}
\end{figure}


\begin{references}

\bibitem{braun1} P. Braun-Munzinger, K. Redlich, J. Stachel, in Quark Gluon Plasma 3, ed. by R.C. Hwa and X.N. Wang (World Scientific Pub., 2004); and references therein.
\bibitem{manninen} J. Manninen and F. Becattini, Phys. Rev. C{\bf 78}:054901 (2008)
\bibitem{rafelski} J. Letessier and J. Rafelski, Eur. Phys. J. A {\bf 35}, 221 (2008)
\bibitem{pbm} A. Andronic, P. Braun-Muzinger and J. Stachel, Phys. Lett. B{\bf 673},142 (2009).
\bibitem{becat} F. Becattini, P. Castorina, A. Milov and H. Satz, Eur. Phys. J. C{\bf 66}. 377 (2010).

\bibitem{star1} J. Adams {\it et al.} (STAR), Nucl. Phys. A{\bf 757}, 102(2005).

\bibitem{star2} J. Adams {\it et al.} (STAR), Phys. Rev. Lett. {\bf 92}, 112301(2004).

\bibitem{phenix} S. S. Adler {\it et al.} (PHENIX), Phys. Rev. C{\bf 69},034909 (2004).


\bibitem{bass} S. A. Bass and A. Dumitru, Phys. Rev. C{\bf 61}, 064909 (2000).

\bibitem{song} H. Song, S.A. Bass and U. Heinz, Phys. Rev. C{\bf 83}, 024912(2011).
\bibitem{hirano} T. Hirano and K. Tsuda, Phys. Rev. C{\bf 66}, 054905 (2002).
\bibitem{gerber} H. Beibe, P. Gerber, J.L. Goity, H. Leutwyler, Nucl. Phys. B{\bf 378} (1992).
\bibitem{teaney} D. Teaney, nucl-th/0204023.


\bibitem{kolb} P.F. Kolb, R. Rapp, Phys. Rev. C{\it 67}, 044903 (2003).
\bibitem{huovinen} P. Huovinen, Eur. Phys. J. A{\bf 37}, 121(2008).



\bibitem{dobler} H. Dobler, J. Sollfrank, and U. Heinz, Phys. Lett. B{\bf457}, 353 (1999).
\bibitem{lee} Kang S. Lee, U. Heinz, and E. Schnedermann, Z.
Phys. {\bf C 48}, 525 (1990).
\bibitem{sollfrank} E. Schnedermann, J. Sollfrank, and U. Heinz, Phys. Rev. {\bf
C 48}, 2462 (1993).

\bibitem{cooper} F. Cooper, G. Frye, Phys. Rev. D {\bf 10}, 186 (1974).
\bibitem{bjorken} J. D. Bjorken, Phys. Rev. D{\bf 27}, 140 (1983).

\bibitem{star3} J. Adams et al.(STAR), Phys. Lett. B{\bf 612}, 181 (2005).
\bibitem{star4} J. Adams et al.(STAR), Phys. Rev. C{\bf 71}, 064902 (2005).
\bibitem{star5} A. Billmeier et al.(STAR), J. Phys. G{\bf 30}, s363(2004).
\bibitem{star6} O. Barannikova et al.(STAR), nucl-ex/0403014.

\bibitem{phobos} B.B. Back {\it et al.}(PHOBOS), Phys. Rev. Lett.{\bf 91}, 052303 (2003).
\end{references}
\end{document}